\documentclass[prl,aps,twocolumn]{revtex4}
\usepackage[dvips]{graphicx}
\usepackage{epsfig}

\begin{document}

\title{Exact solution of a model of time-dependent  evolutionary dynamics
in a rugged fitness landscape  }

\author{Cl\'ement Sire$^1$, Satya N. Majumdar$^2$ and David S. Dean$^1$}

\affiliation{$^1$Laboratoire de Physique Th\'eorique (UMR 5152 du
CNRS), Universit\'e Paul Sabatier, 118, route de Narbonne, 31062
Toulouse Cedex 4, France\\
$^2$Laboratoire de Physique Th\'eorique et Mod\`eles Statistiques (UMR
8626 du CNRS), Universit\'e Paris-Sud, B\^at. 100, 91405 Orsay Cedex,
France.}


\begin{abstract}
A simplified form of the time dependent evolutionary dynamics 
of a quasispecies model with a rugged fitness landscape is solved
via a mapping  onto a random flux model whose asymptotic behavior  
can be described in terms of a random walk.  The statistics of the number 
of changes of the dominant genotype from a finite set of genotypes are 
exactly obtained  confirming existing conjectures based on numerics.    
\end{abstract}

\maketitle 
In  evolution, long periods of stasis or inactivity 
are punctuated by bursts of rapid activity.  
Fossil records \cite{evo} reveal this basic pattern in the evolution of 
biological species and the same behavior is observed in the development of 
microbial populations \cite{micro} and artificial life \cite{life}. 
Not surprisingly, the  dynamics of genetic 
algorithms \cite{algo} also exhibits this punctuated behavior. 
In this paper we will show how a simple model of biological evolution 
can be exactly solved  using a  mapping onto a 
random flux model. The important asymptotic details of this random flux model 
can then be   determined in terms of the first passage  time distribution 
of a random walk.  

The model we study was introduced in \cite{krug} as a simplified
version of the quasispecies model which is used for the study of large
populations of replicating macromolecules \cite{quasi}.  In
\cite{krug}, the quasispecies model was studied in the strong
selection limit where the location in the space of genotypes is
defined as the genotype having the largest population.  A shell model
\cite{krug} may be derived in the strong selection limit and a further
simplification of this model leads to the i.i.d.  (independent and
identically distributed) shell model where the natural space of
genotypes, which is that of binary sequences, is replaced by a one
dimensional lattice. Rather than re-derive the model we shall describe
it and the reader will immediately see that it can be reinterpreted in
terms of a simple evolutionary process.

We consider an ensemble of  $N$ different genotypes labeled
by $i=1,2,\ldots N$. The fitness of a genotype is given by its
effective rate of reproduction per individual $v_i\ge 0$ and thus the
size of the population at time $t$ is given by $n_i(t) =
n_i(0)\exp(v_i t)$. In terms of logarithmic variables, $y_i(t) =
\ln(n_i(t)) = \ln(n_i(0))+ v_i t$. One can interpret $y_i(t)$ as the
trajectory of a particle moving ballistically with a non-negative
velocity $v_i$, starting from its initial position $y_i(0)$.  The
i.i.d. version of the shell model~\cite{krug}, which we will call the
leader model, is defined as follows: we draw $N$ velocities $\lbrace
v_i\rbrace_{1\leq i\leq N}$ independently from the same probability
distribution $p(v)$ (which has positive support only).  We then
consider the semi-infinite lines of slope $v_i$ describing the
evolution of genotype $i$ (up to an overall constant)
\begin{equation}
y_i(t)=-i+v_i\,t.
\end{equation}
At any time $t>0$, the leader is defined as the genotype $i$ having
the maximum $y_i(t)$, the corresponding $i$ is thus the most populated
genotype at time $t$. The choice of $y_i(0) = -i$ comes from the
details of the original quasispecies model \cite{krug}.  Thus, the
evolution of the trajectories is completely deterministic, the only
randomness comes from the velocities.  Obviously at $t=0$, $y_1$ is
the leader; however if $v_1$ is not the maximal velocity, then $y_1$
will ultimately be overtaken by a faster/fitter genotype.  At each of
these overtaking events the number of genotypes which have been leaders
increases by one, finally the fastest genotype will become the final
leader and no more leader changes will occur. In the general context of 
evolutionary processes these over takings correspond to punctuation events.
 
The total number of lead
changes is denoted by $l_N$ and we denote by $w_k$ the velocity of the
leading genotype after the $k$-th lead change. Clearly $l_N$ is a
random variable, varying from one realization of velocities to
another. Based on simulations, it was observed~\cite{krug} that for
large $N$, $\langle l_N\rangle \approx{\beta}\, \ln N$.  where,
remarkably, the coefficient $\beta$ is rather robust and depends only
on the tails of the distribution $p(v)$. Based on numerics, Krug and
Karl~\cite{krug} made some conjectures about the value of $\beta$
and also showed how a comparison with record statistics gives the 
upper bound $\beta <1$.  Similar logarithmic growth of the average 
number of lead changes has also been reported~\cite{krap} 
recently in the context of growing networks where the leader is the 
maximally connected node.

In this letter, we present an exact solution to this problem,
confirming the conjectures of \cite{krug}. Moreover, we calculate the
variance of $l_N$ and show that $\langle (l_N-\langle
l_N\rangle)^2\rangle\approx \gamma \ln N$ for large $N$, where the
coefficient $\gamma$ is calculated exactly and shown to be as robust
as $\beta$.  
We also show that the full distribution of $l_N$ around
its mean is asymptotically Gaussian.  The key observation that leads
to the exact solution of this model is a mapping onto a {\it random
flux} model whose late time properties are identical to those of the
original model. Here, the velocity distribution is chosen as before
but instead of fixing the initial positions $y_i(0)$ of the genotype
$i$ at $-i$, we chose it to be a random variable uniformly distributed
on $[0,-N]$.  From a coarse grained point of view, for a large number
of genotypes, this difference in the initial condition is not expected
to change the asymptotic properties.  In the context of the
quasispecies model, this random initial condition translates to having
the initial population of each genotype having a probability
distribution: ${\rm Prob} (n_i(0)=x)=(xN)^{-1}$, with $\exp(-N)\le
x\le 1$.  An example set of trajectories for $N=4$ and 
where $l_N=2$ is shown in Fig. (1).
\begin{figure}
\psfig{figure=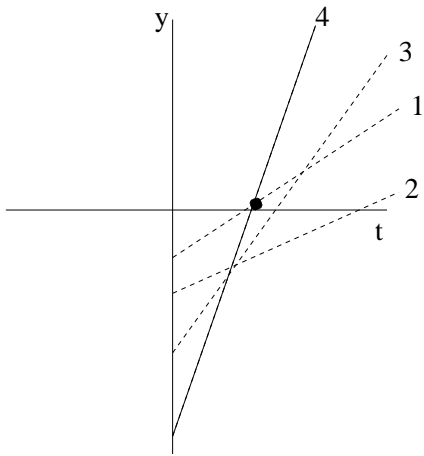,angle=0,height=6cm}
\caption{Model trajectory for $N=4$. Genome $4$ becomes the ultimate leader
and there is only one leadership change indicated by the black dot.}
\label{fig1}
\end{figure}  
If $w_k$ is the velocity of the $k$-th leader then clearly
only genotypes with velocities greater than $w_k$ can become subsequent
leaders.  From the rest frame of the leader, in the next time interval
$\Delta t$ the genotype $i$, with velocity $v_i \ (>w_k) $, will
overtake the leader if it is at a distance $\Delta x = (v_i-w_k)\Delta
t$ behind the leader. The rate at which the genotype $i$ becomes the
new leader is thus given by
\begin{equation}
r_i = (v_i-w_k)\langle \delta\left(y_k(0)- y_i(0)-(v_i-w_k)t\right)\rangle,
\end{equation} 
where the angled brackets indicate the average over the initial conditions.
Given that $y_k(0) > y_i(0)$ the initial distance 
$d_{ik} =y_k(0)- y_i(0) $, 
between the genotypes $i$
and $k$, is a random variable also uniformly distributed 
over  $[0, N+ y_k(0)]$   and consequently the average of the delta function 
in the above  expressions is  equal to one and independent of time. The 
probability that  the genotype $i$ (with $v_i>w_k$) becomes the next 
leader  is  given  by $r_i/\sum_j r_j$  which we  write as a transition 
probability
\begin{equation}
p_{k\to i} =
\frac{(v_i-w_k)\theta(v_i-w_k)}{\sum_{j=1}^N(v_j-w_k)\theta(v_j-w_k)},
\label{tp1}
\end{equation}
This rather intuitive rule appears in a simple traffic model studied
in \cite{kred}, although the physics is different to that here because
on catching up with a slower car the faster one then adopts the same
speed.  We next show that this model can be mapped onto a
first-passage problem for a random process. Notice that, once the
$k$-th leader is selected with velocity $w_k$, the number $N_k$ of
possible future leaders is
\begin{equation}
\frac{N_k}{N}=\frac{\sum_{j=1}^N\theta(v_j-w_k)}{N},
\label{fl1}
\end{equation}
where $N$ is the total number of genotypes. In the limit of large $N$,
one can replace the right hand side of Eq.~(\ref{fl1}) by the integral
over $v$,
\begin{equation}
\frac{N_k}{N}\to \int_{w_k}^{v_{\rm max}} p(v) \,dv = P(w_k),
\label{fl2}
\end{equation}
which is exact up to  $O(1/\sqrt{N})$ corrections and 
where $P(v)=\int_v^{v_{\rm max}} p(u)du $ is the cumulative velocity 
distribution. Clearly, the number of lead changes $l_N$ is the value of 
$k$ where $N_k=1$. This gives, $P(w_{l_N})=1/N$ and hence
\begin{equation}
-\ln[P(w_{l_N})]= \ln N.
\label{level1}
\end{equation}
We  define $Y_k = -\ln[P(w_k)]$ whose evolution is given by
\begin{equation}
Y_{k+1} = Y_{k} + \xi_k,
\label{lange1}
\end{equation}
where clearly
\begin{equation}
\xi_k = -\ln[P(w_{k+1})/P(w_k)].
\label{noise1}
\end{equation}
Thus $Y_k$ can be interpreted as the position of a random walker at
time $k$ and its time evolution is given by the Langevin equation
(\ref{lange1}) where $\xi_k$ is the noise at step $k$. This
redefinition is not yet very useful since the noise at step $k$
depends on $Y_{k+1}$ and $Y_k$. However, as we will see, for large $k$
the probability distribution of the noise $\xi_k$ becomes independent
of $k$ and $w_k$ and has a finite mean $\langle \xi_k \rangle = \mu$
and variance $\langle [\xi_k- \langle \xi_k \rangle]^2 \rangle =
\sigma^2$, that can be computed explicitly for arbitrary velocity
distribution $p(v)$.  For large $k$, Eq.~(\ref{lange1}) represents a
discrete time random walk with a positive drift $\mu$, i.e.,
\begin{equation}
Y_{k+1}= Y_k + \mu + \sigma \eta_k
\label{lange2}
\end{equation}
where $\eta_k$ is a noise with zero mean $\langle \eta_k \rangle =0$
and unit variance. We will also see that $\eta_k$'s are not only 
completely independent of $w_k$ for large $k$, they are also uncorrelated
at different times. Thus Eq.~(\ref{lange2}) is a true Markovian evolution of a
discrete time random walker with a positive drift $\mu$. Obviously then,
by central limit theorem, $Y_k$ will have a Gaussian distribution
with mean $\langle Y_k\rangle= \mu k$ and variance 
${\langle Y_k^2 \rangle-\langle Y_k\rangle}^2 = \sigma^2 k$.

Once we have the Markovian random walker evolution as in Eq.~(\ref{lange2}),
it follows from Eq.~(\ref{level1}) that the number of lead changes 
$l_N$ is just the  first time the process $Y_k$
(starting at some initial value $Y_0$) hits the level $Y=\ln (N)$. 
Thus the distribution of $l_N$ is simply the distribution of the 
first-passage time to the level $Y=\ln (N)$. To compute this, it is 
convenient to define $Z_k=\ln N -Y_k$. Then $Z_k$'s evolve via, 
$Z_{k+1}= Z_k -\mu - \sigma \eta_k$
starting from $Z_0=\ln N -Y_0$. Thus $Z_k$ is the position of a random walker
at step $k$ with a negative drift $-\mu$ towards the origin and $l_N$ now
represents the first-passage time to the {\it origin} starting from the initial
position $Z_0$. Now, for large $k$, the discrete-time random walker 
can be replaced by a continuous-time Brownian motion,
\begin{equation}
\frac{dZ}{dt} = -\mu + \sigma \eta(t)
\label{lange4}
\end{equation}
where $\eta$ is a white noise with $\langle \eta(t)\rangle=0$ and
$\langle \eta(t)\eta(t')\rangle= \delta(t-t')$. For such a process,
the distribution $P(t_f|Z_0)$ of the first-passage time $t_f$ to the
origin is known exactly~\cite{MC} and we can apply it here to obtain the 
probability that $l_N= k$ is given by 
\begin{equation}
Q(k)= \frac{\ln N}{\sigma \sqrt{2\pi k^3}}\, 
\exp\left[-\frac{\mu^2}{2\sigma^2
k}\left(k - (\ln N)/\mu\right)^2\right].
\label{ld}
\end{equation}
Note that this distribution of $l_N$ is non-Gaussian. However, we expect
this result to be valid only in the vicinity of $k\approx \ln N/\mu$, i.e.,
near its mean. This can be traced back to the fact that in deriving
this result we replaced a discrete-time random walk by a continuous-time
Brownian process. Near its mean, using 
$k\approx \ln N$ in Eq.~(\ref{ld}), the distribution of $l_N$ 
becomes a Gaussian
\begin{equation}
Q(k) \approx \frac{\mu^{3/2}}{\sigma\sqrt{2\pi \ln N}} 
\exp\left[-\frac{\mu^3}{2\sigma^2 \ln 
N}\left(k-(\ln N)/\mu\right)^2\right]
\label{gauss1}
\end{equation}
with mean and variance (for large $N$) given by   
\begin{eqnarray}
\langle l_N \rangle &=& \beta\, \ln N; \quad \quad {\rm where}\quad\, \beta=\frac{1}{\mu} 
\label{meanl}\\
\langle (l_N-\langle l_N\rangle)^2 \rangle &=& \gamma\, \ln N; 
\quad\quad {\rm where}\quad\, \gamma=\frac{\sigma^2}{\mu^3}. 
\label{varl}
\end{eqnarray}
Thus, irrespective of the velocity distribution $p(v)$, the distribution of $l_N$ near its mean is
is a universal Gaussian characterized by two parameters 
$\mu$ and $\sigma$. The only dependence on $p(v)$ appears through the two constants $\mu$ 
and $\sigma$.

To calculate the mean $\mu$ and the variance $\sigma^2$ of the noise
$\xi_k$ defined in Eq.~\ref{noise1}, we note that for a given $w_k$,
$\xi_k$ is a random variable since $w_{k+1}$ is a random variable drawn
from the distribution in Eq.~(\ref{tp1}). We define
\begin{eqnarray}
J(v)& = &  \int_v^{v_{\rm max}}P(u)\, du \label{Jv} \\
K(v)&=&  \int_v^{v_{\rm max}} [P'(u)/P(u)] J(u)\, du \label{Kv} \\
L(v) & =&  \int_v^{v_{\rm max}} [P'(u)/P(u)] K(u)\, du \label{Lv}. 
\end{eqnarray}
Using the definition in 
Eq.~(\ref{noise1}) and the transition probability in Eq.~(\ref{tp1}), 
the mean of $\xi_k$ (for a given $w_k$) is 
\begin{equation}
\langle \xi_k \rangle = -\frac{\int_{w_k}^{v_{\rm 
max}}[\ln(P(v))-\ln(P(w_k))](v-w_k)p(v)\,dv}{\int_{w_k}^{v_{\rm max}} (v-w_k) p(v)\, dv}.
\label{mean1}
\end{equation}
Using integration by parts, in both the numerator and denominator above we
find
\begin{equation}
\langle \xi_k \rangle = 1- \frac{K(w_k)}{J(w_k)},
\label{mean2}
\end{equation}
where the function $K(v)$ is defined in Eq.~(\ref{Kv}). The second moment 
is given by
\begin{equation}
\langle \xi_k^2 \rangle = \frac{\int_{w_k}^{v_{\rm
max}}[\ln(P(v))-\ln(P(w_k))]^2(v-w_k)p(v)\, dv}{\int_{w_k}^{v_{\rm max}} (v-w_k) p(v)\, dv},
\label{var1}
\end{equation}
and a similar calculation leads to 
\begin{equation}
\langle (\xi_k-\langle \xi_k\rangle)^2 \rangle = 1+ 2\frac{L(w_k)}{J(w_k)}
-\left[\frac{K(w_k)}{J(w_k)}\right]^2,
\label{var3}
\end{equation}
where the functions $J$, $K$ and $L$ are defined in Eqs. (\ref{Jv}), (\ref{Kv})
and (\ref{Lv}) respectively.

We now consider the three classes of distributions considered by \cite{krug}.

\noindent ({\em i}) {\bf  Fast decaying distribution with $v_{\rm max}=+\infty$}: In 
this case, it is easy to see that for large $u$,
\begin{equation}
\frac{P'(u)}{P(u)} \approx \frac{J'(u)}{J(u)}
\end{equation}
Thus, using this result in the definition of $K(v)$ in Eq.~(\ref{Kv})
one finds that for large $w_k$
\begin{equation}
K(w_k)= \int_{w_k}^{\infty} \frac{P'(u)}{P(u)} J(u)\, du \approx 
 - J(w_k)
\label{KJ}
\end{equation}
Similarly, for large $w_k$, 
\begin{equation}
L(w_k)= \int_{w_k}^{\infty} \frac{P'(u)}{P(u)} K(u) \, du \approx 
J(w_k)
\label{LJ}
\end{equation}
Using these results in Eqs. (\ref{mean2}) and (\ref{var3}) 
we find for large $k$
\begin{eqnarray}
\langle \xi_k \rangle & =\mu=& 2 \label{c1mean}\\
\langle (\xi_k-\langle \xi_k\rangle)^2 \rangle & = \sigma^2=& 2 \label{c1var}.
\end{eqnarray}
Thus, as stated earlier, we see the variance become independent of $k$
and $w_k$.

\noindent{\em (ii)} {\bf  Distribution with a finite $v_{\rm max}$, with $p(v)\sim
|\ln(v_{\rm max}-v)|^\gamma(v_{\rm max}-v)^\alpha$:} In this case, 
for $u$ close to $v_{\rm max}$, we find
\begin{equation}
\frac{P'(u)}{P(u)}\approx \left(\frac{1+\alpha}{2+\alpha}\right)\, \frac{J'(u)}{J(u)}.
\label{edge1}
\end{equation}
and it follows 
that for $w_k$ close to $v_{\rm max}$
\begin{eqnarray}
K(w_k) &\approx & -\left(\frac{1+\alpha}{2+\alpha}\right)\, J(w_k) \nonumber \\
L(w_k) & \approx & \left(\frac{1+\alpha}{2+\alpha}\right)^2 J(w_k) \label{KL}
\end{eqnarray}
Using these results in Eqs. (\ref{mean2}) and (\ref{var3}) we get
\begin{eqnarray}
\langle \xi_k \rangle & =\mu=& \frac{2\alpha+3}{\alpha+2} \label{c2mean}\\
\langle (\xi_k-\langle \xi_k\rangle)^2 \rangle & = \sigma^2=& 
\frac{2\alpha^2+6\alpha+5}{(\alpha+2)^2} \label{c2var}.
\end{eqnarray}

\noindent {\em (iii)} {\bf  Power-law decaying distribution with $v_{\rm max}=+\infty$, and
$p(v)\sim \ln(v)^\gamma v^{-\alpha}$ with $\alpha>2$:} In this case, for large $u$
\begin{equation}
\frac{P'(u)}{P(u)}\approx \left(\frac{\alpha-1}{\alpha-2}\right)\, \frac{J'(u)}{J(u)}
\label{edge2}
\end{equation}
Using this result in the definition of $K(v)$ and $L(v)$ one easily finds
that for large $w_k$ 
\begin{eqnarray}
K(w_k) &\approx & -\left(\frac{\alpha-1}{\alpha-2}\right)\, J(w_k) \nonumber \\
L(w_k) & \approx & \left(\frac{\alpha-1}{\alpha-2}\right)^2 J(w_k) \label{KL2}
\end{eqnarray}
Using these results in Eqs. (\ref{mean2}) and (\ref{var3}) we get
\begin{eqnarray}
\langle \xi_k \rangle & =\mu=& \frac{2\alpha-3}{\alpha-2} \label{c3mean}\\
\langle (\xi_k-\langle \xi_k\rangle)^2 \rangle & = \sigma^2=&
\frac{2\alpha^2-6\alpha+5}{(\alpha-2)^2} \label{c3var}.
\end{eqnarray}

One can also demonstrate \cite{inprep} that for all these velocity
distributions, and for large $k$ and $k'$ $\langle \xi_k \xi_k'\rangle
-\mu^2 \to 0$, indicating that the noise $\xi_k$'s become completely
uncorrelated in time.  Thus Eq.~(\ref{lange2}) truly represents a
Markovian random walk with drift $\mu$.  Knowing the exact values of
$\mu$ and $\sigma$, we then find that distribution of $l_N$, near its
mean, is given by the Gaussian in Eq.~(\ref{gauss1}) with mean and
variance given by Eqs.~(\ref{varl}). The coefficients $\beta$ and
$\gamma$ are thus calculated exactly knowing $\mu$ and $\sigma$ and
are given, for each of the cases mentioned above, by
\begin{eqnarray} (i) &:&\beta =  1/2 \ ; \ \gamma =  1/4 \\
(ii) &:&\beta  =\frac{\alpha+2}{2\alpha+3}  
\ ; \ \gamma= \frac{(\alpha+2)(2\alpha^2+6\alpha+5)}{(2\alpha+3)^3} \\
(iii) &:&\beta= \frac{\alpha-2}{2\alpha-3}\ ; 
\ \gamma  =  \frac{(\alpha-2)(2\alpha^2-6\alpha+5)}{(2\alpha-3)^3} 
\end{eqnarray}

The results for the coefficient $\beta$ are in complete agreement with 
those conjectured in \cite{krug} in all three cases and we have further 
verified all our results  by simulating the original i.i.d. shell model 
with an  algorithm which permits us to simulate up to $N = 10^{200}$ 
genotypes \cite{inprep}. Moreover, we have also calculated the variance
exactly and shown that near its mean, the distribution of $l_N$ 
is a universal Gaussian. In \cite{krug} it was pointed ou that the variance 
of $l_N$ is typically smaller than the mean indicating the  temporal
correlation  between leadership changes, this is clearly seen in our 
exact results.  Away from its mean, one expects
to see departures of the distribution of $l_N$ away from the Gaussian form. 
To compute the full distribution one needs to solve the first-passage 
problem for the discrete-time process without resorting to the
continuous-time approximation. Fortunately, for our discrete-time process,
this can be achieved by observing that the
the evolution of $Y_k$ with $k$, though random, is actually a strictly 
monotonic process. This follows from Eq.~(\ref{noise1}) that shows that
the noise $\xi_k$ is always positive. The distribution
of the first-passage time $l_N$ to
the level $\ln(N)$ then satisfies the identity~\cite{inprep}
\begin{equation}
{\rm Prob}(l_N \leq k) = {\rm Prob}(Y_k \geq \ln(N)).
\end{equation} 
This gives 
$Q(k) = {\rm Prob}(l_N = k) = {\rm Prob}(l_N \leq k+1) -{\rm Prob(l_N \leq k)}
=  {\rm Prob}(Y_{k+1} \geq \ln(N)) - {\rm Prob}(Y_{k} \geq \ln(N))$.
Thus, a knowledge of the distribution of $Y_k$ (which
is usually much simpler to compute) provides us with
an exact distribution of lead changes $Q(k)$ for all $k$. 
For example,   
for an exponential velocity distribution $p(v)=e^{-v}$, 
the probability density function of $Y_k$ can 
be found explicitly for all $k$ 
\begin{equation}
\rho_k(y) = {y^{2k-1}\over (2k-1)!}\exp(-y).
\end{equation}
This result is in fact asymptotically valid for any rapidly decaying
distribution $p(v)$ \cite{inprep}.  Using this result we thus obtain
the full probability distribution of $l_N$ for the exponential
velocity distribution as
\begin{equation}
Q(k) = {(\ln(N))^{2k}\over N (2k)!}\left[ 1 + {\ln(N)\over 2k+1}\right]
\label{eqexp}
\end{equation}
In Fig. (2) we show the predictions of Eq.~(\ref{eqexp}) versus the results
of extensive simulations and the agreement is perfect. The above use of the 
monotonicity of $Y_k$ also enables one to obtain analytical results, 
away from the Gaussian regime, for generic  
fitness distributions \cite{inprep}. 
\begin{figure}
\psfig{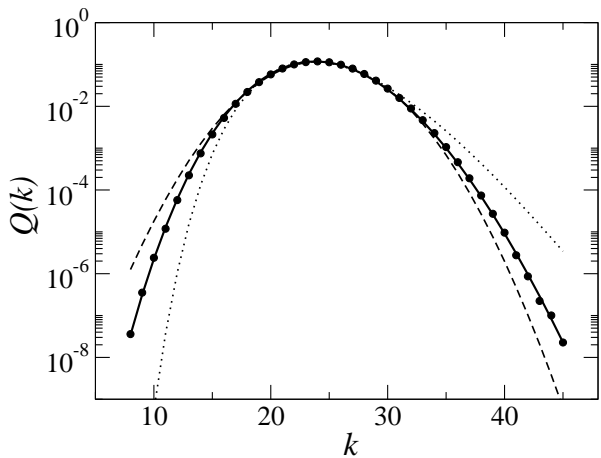}
\caption{Plot of the distribution $Q(k)$ of and $l_N$ (circles), for
$N=10^{20}$ generated from $2.10^8$ samples with velocities taken from
an exponential distribution. Also shown is the result
Eq.~(\ref{eqexp}) (solid line), the result Eq.~(\ref{ld}) (dotted
lines) and the Gaussian result Eq.~(\ref{gauss1}) (dashed line).}
\label{fig2}
\end{figure}  

To summarize we have solved exactly the asymptotic statistics of lead
changes in a quasispecies evolution model by mapping the model to a
random flux model.  Our results confirm previous conjectures about the
mean number of leader changes. We have also computed the variance
exactly and shown that the distribution is generically Gaussian in the
region around the mean. Finally, we remark that the evolution time $\tau$
defined as the time when the last leader change occurs can be shown to 
have a distribution $q(\tau)\sim \tau^{-2}$ for large $\tau$ \cite{inprep}, 
as found in more realistic models \cite{krug}. 

\noindent{\bf Acknowledgment}: SNM and DSD would like to thank 
the Isaac Newton Insitute Cambridge where part of this work was 
carried out during the program {\em  Principles of the  
Dynamics of Non-Equilibrium Systems}. The authors would also
like to thank J. Krug for useful comments and suggestions.

\end{document}